\documentclass[sigconf,natbib=true,anonymous=false]{acmart}

\renewcommand\footnotetextcopyrightpermission[1]{}
\AtBeginDocument{%
  }

\begin{document}

\title{Faithful or Findable? Evaluating LLM-Generated Metadata for RDF Dataset Search}

\thispagestyle{plain}
\pagestyle{plain}

\author{Riccardo Terrenzi}
\orcid{1234-5678-9012}
\email{rite@mmmi.sdu.dk}
\affiliation{%
  \institution{University of Southern Denmark}
  \city{Sønderborg}
  \country{Denmark}
}

\author{Serkan Ayvaz}
\affiliation{%
  \institution{University of Southern Denmark}
  \city{Sønderborg}
  \country{Denmark}
}
\email{seay@mmmi.sdu.dk}


\begin{abstract}
Dataset search depends heavily on metadata, making LLM-generated metadata a consequential form of synthetic content in retrieval systems. We study six metadata-generation settings for RDF datasets, ranging from simple rewriting to profile-grounded and agentic graph-based generation, and evaluate them jointly for retrieval effectiveness and faithfulness. Unconstrained metadata rewriting delivers the strongest retrieval gains over the original metadata, but it is also the least faithful, showing that search improvements can be driven by unsupported semantic expansion. More grounded settings substantially improve faithfulness, and profile-grounded rewriting provides the most balanced trade-off between retrieval effectiveness and grounding. These findings position synthetic metadata as a system-level IR problem in which effectiveness, provenance, and trust must be evaluated together.
\end{abstract}




\maketitle

\section{Introduction}
Dataset search depends heavily on metadata~\cite{chapman2020dataset,paton2024dataset}. Unlike many traditional document retrieval settings, users rarely interact directly with the underlying data at search time; instead, retrieval systems index and rank dataset surrogates such as titles, descriptions, and keywords~\cite{brickley2019googledatasetsearch}. This makes metadata a particularly consequential layer in the retrieval pipeline: changing metadata can change how datasets are represented, matched, and ultimately discovered. Recent advances in large language models (LLMs) and tool-using agents make it increasingly feasible to rewrite, enrich, or generate dataset metadata automatically, raising the prospect of synthetic metadata becoming part of real-world search infrastructures~\cite{zhang2025autoddg}.

From an information retrieval perspective, this shift is promising but non-trivial. On the one hand, automated metadata generation may improve sparse or incomplete descriptions and expose signals that help ranking models retrieve datasets more effectively~\cite{chen2024snippets,zhang2025autoddg}. On the other hand, synthetic metadata is not a neutral transformation: it can introduce new terms, new levels of specificity, and new interpretations of the underlying dataset. In mixed human--synthetic retrieval environments, this creates a central question not only about effectiveness, but also about grounding. If synthetic metadata improves search performance while drifting away from the original metadata or the dataset itself, then the retrieval gains come with corresponding risks for faithfulness, provenance, and trust.

In this paper, we investigate this problem on a subset of ACORDAR 2.0~\cite{chen2024acordar2} containing roughly 1,000 RDF datasets. We study several metadata production settings that reflect increasingly automated forms of synthesis: \texttt{metadata rewrite} from original metadata, \texttt{profile rewrite} aided by a dataset profile, \texttt{profile gen} from profile-based inputs, \texttt{profile title gen} that retains the original title, and agentic generation with direct access to the underlying dataset through \texttt{dataset agent} and \texttt{dataset title agent}. We evaluate the resulting metadata from two complementary perspectives. First, we benchmark their impact on dataset retrieval using standard IR metrics, including NDCG, MAP, and MRR. Second, we assess faithfulness through an LLM-based judge, with claim-level verification against the appropriate evidence source for each setting, including original metadata, profiles, and the underlying dataset.

Our aim is therefore not simply to determine whether synthetic metadata works, but to understand how synthetic metadata changes the retrieval representation of datasets and what faithfulness costs may accompany those changes. At this stage, we take a cautious and exploratory view: synthetic metadata and automated generation pipelines appear increasingly plausible for deployment, but their effect on search should be understood together with their effect on grounding. By examining both retrieval outcomes and faithfulness behavior across multiple generation settings, we position dataset search as a concrete and consequential case study for broader questions about synthetic content in IR systems.

\section{Related Work}

Dataset search retrieves over dataset surrogates such as titles, descriptions, keywords, and other metadata rather than raw data, making metadata quality central to effectiveness. This view is established in dataset search research~\cite{chapman2020dataset,paton2024dataset} and systems such as Google Dataset Search~\cite{brickley2019googledatasetsearch}. For RDF dataset retrieval, benchmarks including ACORDAR~\cite{lin2022acordar} and ACORDAR~2.0~\cite{chen2024acordar2} enable systematic evaluation.

Prior work has studied better dataset representations for search. Compact data snippets improve retrieval by surfacing salient information missing from sparse metadata~\cite{chen2024snippets}, and later work extends this with multi-objective optimization over relevance, compactness, and representativeness~\cite{zhou2025mds}. More broadly, dataset discovery in data lakes likewise highlights the importance of representation and indexing choices~\cite{bogatu2020datalakes}.

Large language models have made automatic dataset description generation more practical. AutoDDG is especially relevant because it frames description generation as a discovery-oriented task and evaluates LLM-generated descriptions for search utility~\cite{zhang2025autoddg}. Our work is closest to this line, but asks not only whether generated metadata improves retrieval, but also whether it remains faithful to available evidence. We use BM25~\cite{robertson2009bm25} as a standard lexical baseline.

Our work is also informed by metadata quality and RDF profiling. Studies of open data portals show that metadata is often incomplete, inconsistent, or poorly aligned, harming discoverability~\cite{neumaier2016qualityassessment,noguerasiso2021metadataquality,kubler2018metadataquality,bogdanovic2023alignment,kremen2019discoverability}. In parallel, RDF profiling provides principled graph summaries~\cite{ellefi2018rdfprofiling}. Systems such as LODStats~\cite{auer2012lodstats,ermilov2016lodstats} and Description Set Profiles~\cite{honma2014dsp} show that structural and semantic properties of RDF datasets can be extracted automatically, making profiling a natural bridge between raw RDF data and retrieval-oriented synthetic metadata.

\section{Synthetic Metadata Generation}

We construct different metadata settings to test how different forms of grounding affect the generation of search-oriented dataset representations. The central methodological question is whether LLMs should operate only on the original metadata, on a compact dataset-level abstraction of the RDF source, or by directly exploring the RDF graph. We therefore design a small family of generation settings that differ only in the evidence made available to the model and in the degree of control imposed on the output.

For all RDF-grounded settings, we first transform each local RDF source into a standardized dataset profile~\cite{ellefi2018rdfprofiling,auer2012lodstats,honma2014dsp}. Rather than exposing raw triples or arbitrary instance samples, the profile aggregates recurring signals at dataset level: graph statistics, dominant namespaces and types, per-type property profiles with coverage and multiplicity information, frequent typed relations, lexical cues from labels and URI local names, and scope indicators such as domain, geography, and time when supported by the graph~\cite{ellefi2018rdfprofiling,auer2012lodstats,honma2014dsp}. A small number of representative records is included only to anchor the summary in concrete evidence. This design is motivated by two constraints: it keeps the prompt compact enough for efficient generation, and it reduces the risk that metadata is driven by distinctive rows instead of stable structural regularities.

On top of this profile, we define three prompt-based settings. In \texttt{metadata rewrite}, the model receives only the original metadata and rewrites it to improve lexical findability while preserving the original schema. This setting isolates the effect of search-oriented rewriting without RDF grounding. In \texttt{profile rewrite}, the model receives both the original metadata and the dataset profile, and again produces a rewritten full record. This setting tests whether RDF-derived evidence can enrich an existing metadata description while retaining its original structure. In \texttt{profile gen}, the model receives only the dataset profile and generates a new search metadata record from scratch, restricted to \texttt{title}, \texttt{description}, and \texttt{tags}. This condition is intentionally stricter: it measures how much of the searchable representation can be recovered from RDF evidence alone, without inheriting wording from the original metadata. We also evaluate \texttt{profile title gen}, a controlled ablation variant in which the original title is provided as optional, non-authoritative context. The motivation is to test whether a minimal lexical hint can stabilize generation when the RDF content is structurally informative but weak in surface cues~\cite{chen2024snippets}.

Finally, we evaluate an agentic setting in which the model does not rely on a precomputed summary, but interacts directly with the local RDF graph through read-only tools for graph overview, metadata-candidate inspection, triple search, neighborhood exploration, and restricted SPARQL queries. As in \texttt{profile gen}, the agent generates only \texttt{title}, \texttt{description}, and \texttt{tags}, which keeps the comparison focused on the search-critical fields. We consider both a fully graph-grounded variant, \texttt{dataset agent}, and a title-context variant in which the original title is available only as a non-authoritative hint. This final comparison allows us to separate the value of direct graph exploration from the value of profile-based compression.

\section{Search Performance Evaluation}

We evaluated search performance on the ACORDAR benchmark~\cite{chen2024acordar2}, which supports the full retrieval loop needed for this study: a corpus of dataset metadata records, a topic set, and relevance judgments (\texttt{qrels}). This setup made it possible to measure the effect of metadata replacement directly at retrieval time, rather than assessing rewritten metadata only qualitatively.

Because full-scale evaluation over ACORDAR was computationally expensive, we constructed a budgeted pilot subset under a hard cap of 1{,}000 indexed documents while preserving heterogeneity in query difficulty and relevance density. The subset was derived from the full ACORDAR metadata, topics, and \texttt{qrels}. A topic was considered eligible only if it had at least three positive relevance judgments and a complete baseline ranking through rank~20.

Topic selection was designed to balance diversity and indexing cost. We stratified eligible topics using two signals: baseline BM25~\cite{robertson2009bm25} \texttt{nDCG@10}, used as a proxy for difficulty and divided into tertiles (\texttt{hard}, \texttt{medium}, \texttt{easy}), and the number of positive \texttt{qrels}, divided into tertiles (\texttt{sparse}, \texttt{medium}, \texttt{dense}). We then applied a stratified round-robin policy over the resulting nine strata. At each step, the selected topic minimized incremental document cost, defined as the number of previously unseen documents required to include both its judged documents and its top-20 baseline documents. The final pilot contained 37 topics and 996 documents, covering all nine strata. Within this fixed pilot, we compared all previously introduced metadata-generation settings against a common \texttt{original} baseline using the unmodified metadata. The main prompt-based runs used \texttt{openai/gpt-5.4-mini}, while the agentic runs used \texttt{mistralai/mistral-small-2603}.

All settings were evaluated with two retrieval backends under otherwise fixed conditions. For BM25~\cite{robertson2009bm25}, each run indexed the same 996 pilot documents and retrieved up to \texttt{top k = 100} results per query. The indexed representation used a weighted tokenization scheme in which title tokens were weighted by~3, tags by~2, and category by~2, together with normalized tokens from the remaining metadata fields. For dense retrieval, each run used the same pilot documents and retrieval depth, but represented each dataset as a concatenation of labeled metadata fields, embedded with \texttt{BAAI/bge-base-en-v1.5}, and retrieved with cosine similarity in a persistent vector database. Retrieval backends were not tuned per setting; the only varying factor was the metadata itself.

For both backends, we report MAP, NDCG@10, MRR@10, Precision@10, and Recall@10. Among these, NDCG@10 was treated as the primary ranking metric because it emphasizes the top of the ranking while preserving graded relevance. Each candidate setting was compared against the same \texttt{original} baseline, and we computed both aggregate metric deltas and topic-level win/loss/tie counts on NDCG@10. Since all 37 topics in the final pilot had positive \texttt{qrels}, all averages were computed over the same topic set in every run.

\begin{table}[t]
\centering
\footnotesize
\setlength{\tabcolsep}{4pt}
\renewcommand{\arraystretch}{0.98}
\caption{Retrieval effectiveness of the original and generated metadata variants.}
\label{tab:retrieval-results}
\begin{tabular*}{\columnwidth}{@{\extracolsep{\fill}}lcccc@{}}
\toprule
Method & MAP & nDCG@10 & MRR@10 & R@10 \\
\midrule
\multicolumn{5}{@{}l}{\textbf{BM25}} \\
\midrule
Original metadata        & 0.4520 & 0.4487 & 0.6867 & 0.3338 \\
Metadata Rewrite      & 0.4562 & 0.4782 & 0.7951 & 0.3220 \\
Profile Rewrite       & 0.4482 & 0.4644 & 0.7489 & 0.3330 \\
Profile Gen           & 0.2555 & 0.3356 & 0.6273 & 0.2413 \\
\textit{+ title}     & 0.3809 & 0.4324 & 0.7279 & 0.3157 \\
Dataset Agent         & 0.2802 & 0.3494 & 0.6153 & 0.2513 \\
\textit{+ title}   & 0.3471 & 0.4105 & 0.7060 & 0.2761 \\
\midrule
\multicolumn{5}{@{}l}{\textbf{Dense}} \\
\midrule
Original metadata        & 0.5144 & 0.5628 & 0.7833 & 0.3990 \\
Metadata Rewrite      & 0.5131 & 0.5766 & 0.7680 & 0.4004 \\
Profile Rewrite       & 0.5031 & 0.5652 & 0.7270 & 0.4028 \\
Profile Gen           & 0.3038 & 0.3740 & 0.6314 & 0.2858 \\
\textit{+ title}     & 0.4251 & 0.4878 & 0.8207 & 0.3582 \\
Dataset Agent         & 0.3678 & 0.4180 & 0.7721 & 0.3061 \\
\textit{+ title}   & 0.4551 & 0.4982 & 0.7646 & 0.3838 \\
\bottomrule
\end{tabular*}
\end{table}

Table~\ref{tab:retrieval-results} highlights a consistent pattern across retrieval backends. \texttt{metadata rewrite} was the strongest prompt-based condition, whereas scratch-style generation generally reduced effectiveness relative to the baseline. Providing the title mitigated some of this loss for both profile-based and agentic generation, but these grounded variants still did not reach the performance of \texttt{metadata rewrite}. Overall, dense retrieval remained stronger than BM25 in absolute terms across all settings.

\section{Faithfulness Evaluation}

We evaluate faithfulness as a claim-level judgment task over six metadata-generation settings. Because verifying every generated statement over the full pilot benchmark would be impractical, we conduct the analysis on three disjoint subsets sampled from the ACORDAR pilot collection~\cite{chen2024acordar2}, for a total of 150 datasets. To preserve coverage across RDF sources of different structural complexity, subset construction is stratified by the number of distinct predicates in the dataset profile, which we use as a proxy for semantic complexity~\cite{ellefi2018rdfprofiling,auer2012lodstats,ermilov2016lodstats}. We define low-, medium-, and high-complexity strata and balance each subset across them, yielding three comparable evaluation sets while keeping the annotation workload tractable.

Faithfulness is not measured directly on whole metadata strings. Instead, for each generated metadata, we first extract atomic claims using an LLM and then judge each claim against the evidence that is authoritative for the corresponding setting. In particular, an LLM receives each metadata (one field per time) and extracts atomic claims. Claim extraction is intentionally conservative: claims must be explicit or very directly implied by the generated text, compound statements are split into atomic units, and duplicates are removed. The extracted claims are typed into a small set of semantic categories, including topic, content, granularity, location, time, dataset type, provenance, and use case. This representation allows us to evaluate the generated metadata at the level of individual semantic assertions rather than overall fluency.

The six settings differ in what evidence is considered authoritative. In \texttt{metadata rewrite}, claims are judged against the original metadata only. In \texttt{profile rewrite}, they are judged against the original metadata together with the RDF-derived dataset profile~\cite{ellefi2018rdfprofiling,honma2014dsp}. In \texttt{profile gen} and \texttt{profile title gen}, the evidence consists of the profile alone or the profile plus the original title, respectively. In \texttt{dataset agent} and \texttt{dataset title agent}, claims are judged against the full RDF graph, again with or without the original title as auxiliary context. This design allows us to compare faithfulness under progressively stronger forms of grounding, from metadata-only evidence to direct access to the underlying dataset.

The evaluation procedure itself also differs across these settings. For \texttt{metadata rewrite}, \texttt{profile rewrite}, \texttt{profile gen} and its ablation, the evidence fits into a compact textual input, so we use a standard LLM judge. For each dataset, the judge receives the relevant evidence together with the extracted claims, and returns one faithfulness label per claim. In contrast, for \texttt{dataset agent} and its ablation, 
we use an RDF-grounded agentic judge that interacts directly with the local graph through read-only tools for graph overview, triple search, resource inspection, and restricted SPARQL querying. The agent is instructed to examine the graph, verify each claim against local evidence, and return the same structured claim-level labels as in the non-agentic settings. This distinction is methodologically important: the simpler settings test whether claims are supported by compact evidence representations, whereas the dataset-level settings test whether claims remain faithful when judged against the full source graph.

\begin{figure}[t]
  \centering
  \includegraphics[width=\columnwidth]{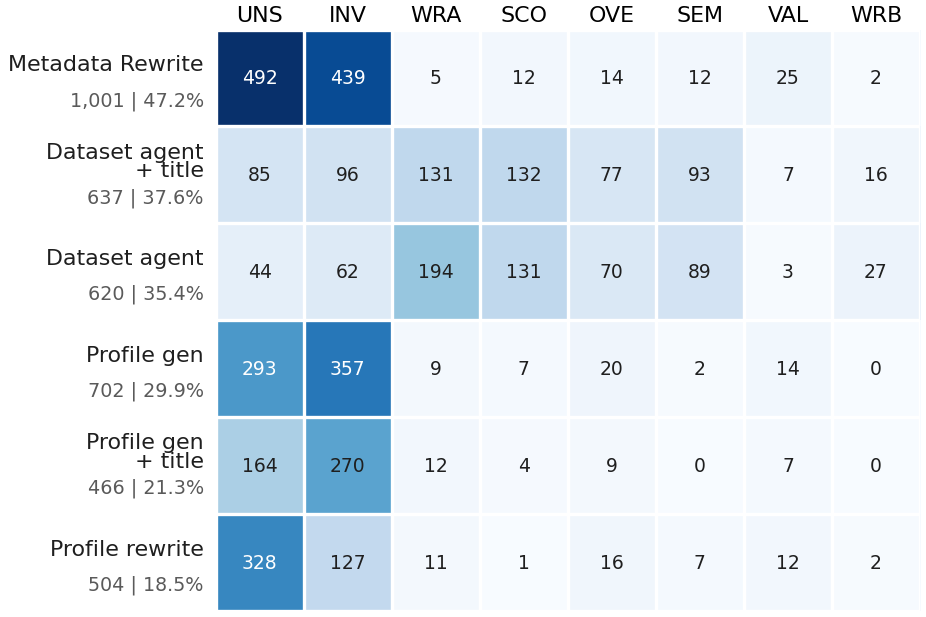}
  \caption{Failure modes across the different metadata generation settings (UNS = Unsupported Inference, INV = Invented Detail, WRA = Wrong Entity/Type, SCO = Scope Mismatch, OVE = Overgeneralization, SEM = Semantic Drift, VAL = Value Judgement, WRB = Wrong Relation).}
  \Description{Visualization of failure modes for metadata generation settings.}
  \label{fig:heatmap}
\end{figure}

Each claim is assigned one of four labels: supported, partially supported, contradicted, or insufficient evidence. We can see the aggregate faithfulness results in Table~\ref{tab:faithfulness-results}.
For non-supported claims, we additionally record a single failure reason, including overgeneralization, unsupported inference, invented detail, wrong entity or relation, scope mismatch, value judgment, or semantic drift. Results are aggregated at the claim level. Our primary measure is the supported-claim rate, complemented by the rates of partially supported, contradicted, and insufficient-evidence claims, as well as aggregate analyses of the dominant failure modes, as shown in Fig.~\ref{fig:heatmap}. 
Tables~\ref{tab:faithfulness-results} and Fig.~\ref{fig:heatmap} indicate that \texttt{metadata rewrite} is by far the least faithful setting, whereas, as expected, more grounded settings achieve substantially better faithfulness. In addition, the agentic setting underperforms relative to what one might expect. Across settings, the most frequent types of unfaithful claims are Unsupported Inference and Invented Detail as expected.

\begin{table}[t]
\centering
\footnotesize
\setlength{\tabcolsep}{3.5pt}
\renewcommand{\arraystretch}{0.98}
\caption{Aggregate faithfulness results across the three evaluation subsets. Values are claim-level rates.}
\label{tab:faithfulness-results}
\begin{tabular*}{\columnwidth}{@{\extracolsep{\fill}}lcccccc@{}}
\toprule
Setting & Supported & Partially & Contradicted & Insuf. Evidence \\
\midrule
Metadata Rewrite    & 0.5283 & 0.0858 & 0.0523 & 0.3336  \\
Profile Rewrite     & 0.8152 & 0.0550 & 0.0059 & 0.1239  \\
Profile Gen         & 0.7005 & 0.0525 & 0.0064 & 0.2406  \\
\textit{+ title}   & 0.7872 & 0.0502 & 0.0151 & 0.1475  \\
Dataset Agent       & 0.6459 & 0.0965 & 0.0200 & 0.2376  \\
\textit{+ title} & 0.6237 & 0.0910 & 0.0372 & 0.2481 \\
\bottomrule
\end{tabular*}
\end{table}

\section{Discussion}

Our results show that LLM-based metadata generation can improve retrieval effectiveness over the original human-authored metadata, but that these gains conceal an important validity issue. The best retrieval performance is obtained by the unconstrained \texttt{metadata rewrite} variant, which improves over the original metadata on both BM25 and dense retrieval. However, the same method is also the least faithful. Fig.~\ref{fig:heatmap} suggests that part of the retrieval gain is achieved by introducing unsupported inferences and invented details that are useful for matching queries, but are not fully supported by the source metadata. In other words, stronger retrieval does not necessarily indicate better metadata quality: it may instead reflect a shift toward synthetic metadata that is more search-effective but less reliable.

This raises a central question: are the observed errors primarily caused by the limited informativeness of the original metadata, or by insufficient grounding in the generation process? Our experiments suggest that both factors matter. On the one hand, the success of \texttt{metadata rewrite} indicates that the original metadata are often too sparse or insufficiently query-aligned for effective retrieval~\cite{chapman2020dataset,noguerasiso2021metadataquality,kremen2019discoverability}. On the other hand, the weak faithfulness of \texttt{metadata rewrite} shows that unconstrained generation is prone to unsupported inference and invented detail. The challenge is therefore not simply to add more text, but to generate metadata that is simultaneously informative for retrieval and faithful to the underlying data source.

The profile-based experiments help isolate this tradeoff. When the LLM is provided with a dataset profile, faithfulness improves substantially, especially if the original title is provided as well, but retrieval remains below the original baseline. This indicates that grounding alone is not enough: the profile must also be sufficiently informative to support retrieval-oriented metadata generation. The current results therefore point to an open problem in RDF dataset summarization~\cite{ellefi2018rdfprofiling,auer2012lodstats,honma2014dsp}. Existing profiles appear to preserve faithfulness, but they do not yet capture enough of the semantic and lexical content needed for strong search performance. A key direction for future work is to develop profile extraction methods that preserve semantic fidelity while also surfacing the aspects of the dataset that are most useful for retrieval.

The agentic setting offers a complementary perspective. If a compact profile is not informative enough, a natural alternative is to allow the model direct access to the full dataset. This approach is promising because it removes the bottleneck imposed by profile quality and creates the possibility of richer, more context-sensitive metadata generation. In our results, however, \texttt{dataset agent} does not yet match the retrieval performance of the best rewrite-based approach, and its faithfulness is still not strong enough to resolve the problem. Even with access to the original dataset, the generation process remains vulnerable to abstraction errors such as entity/type mismatch and scope mismatch. This suggests that source access alone is insufficient and gives a lead for future works about metadata-generating agents equipped with stronger mechanisms for evidence selection, grounding control, and provenance-aware abstraction.

Overall, the results support a broader conclusion about synthetic content in retrieval systems. Metadata generation should not be evaluated only in terms of search effectiveness, because retrieval gains may be partially driven by unfaithful content. Instead, generated metadata should be assessed as a form of synthetic content in a mixed human--AI retrieval ecosystem, where utility, provenance, and faithfulness must all be considered jointly~\cite{zhang2025autoddg,chen2024snippets}. In this sense, \texttt{profile rewrite} is the most encouraging current result: it does not maximize retrieval, but it provides the strongest balance between retrieval effectiveness and faithfulness.

\section{Conclusion}

This paper examined synthetic metadata generation for dataset search as both an effectiveness problem and a grounding problem. Across multiple generation settings, we found that LLM-based metadata rewriting can improve retrieval performance over the original human-authored metadata, but that these gains do not necessarily correspond to better metadata quality. In particular, the strongest retrieval improvements came from unconstrained rewriting, which also produced the weakest faithfulness results, showing that search effectiveness can be partly driven by unsupported semantic expansion rather than by more accurate dataset representation. At the same time, more grounded approaches based on dataset profiles or direct RDF access improved faithfulness, but did not yet match the retrieval strength of the best rewrite-based setting. Among the tested methods, profile-grounded rewriting offered the most convincing balance between retrieval utility and semantic grounding. Overall, our findings suggest that synthetic metadata should be treated as a form of synthetic content in IR systems, where effectiveness, provenance, and trust must be evaluated jointly. Future work should focus on richer RDF profiling methods, stronger grounding controls for agentic generation, and evaluation protocols that better capture the trade-off between retrieval gains and metadata faithfulness.

\bibliographystyle{ACM-Reference-Format}
\bibliography{bibliography}

\end{document}